\documentclass{jltp}
\include{psfig}

\title{Current-Voltage Relations in $d$-wave Josephson Junctions: Effects of Midgap Interface States}

\author{T. L\"ofwander, G. Johansson, and G. Wendin}

\address{Dep. of Microelectronics and Nanoscience, School of Physics and Engineering\\
Physics, Chalmers University of Technology and G\"oteborg University,\\
S-412 96 G\"oteborg, Sweden\\}

\runninghead{T. L\"ofwander, G. Johansson, and G. Wendin}{Current-Voltage Relations in $d$-wave Josephson Junctions...}

\begin{document}

\maketitle

\begin{abstract}
  We investigate the dc current-voltage characteristics of $d$-wave
  Josephson junctions, where the barrier at the interface may have
  arbitrary strength. Dividing the current into $n$-particle currents
  $I_n$ ($n$ integer), we can explicitly show which physical processes
  are responsible for the subharmonic gap structure (SGS). For
  orientations where midgap states (MGS) exist, the resonances in the
  $n$-particle processes are drastically changed, giving rise to a
  strongly modified SGS.  Introducing broadening in a phenomenological
  way, we show that MGS may produce a current peak near zero bias and
  we explain which physical processes are contributing to this peak.
  The agreement of our theory with recent experiments is discussed.

\end{abstract}

\section{INTRODUCTION}
The formation of midgap states\cite{Hu} (MGS) at surfaces and
interfaces of $d$-wave superconductors affects the current transport
properties of junctions involving $d$-wave superconductors. It has
been established\cite{Cov,Alff_NS} that the zero-bias conductance peak
(ZBCP) seen in normal metal/$d$-wave superconductor (N/$d$) junctions,
are due to the MGS. For the ac Josephson effect it was
shown\cite{Hurd,BarSvi} that subharmonic gap structure (SGS) is in
general different in $d$-wave junctions compared to $s$-wave
junctions. In Ref.~\onlinecite{Hurd} it was shown that MGS produce
current peaks at voltages of the order of the maximum gap, while in
Ref.~\onlinecite{BarSvi} also a singularity at zero bias was found.
Later on, in Ref.~\onlinecite{SamDat}, it was pointed out that this
structure was not seen in Ref.~\onlinecite{Hurd}, because the
scattering theory did not include broadening effects. Numerical
calculations\cite{Yoshida} including broadening confirmed the picture
outlined in Ref.~\onlinecite{SamDat}.  Despite all the interest in the
ac Josephson effect for $d$-wave junctions, the physics behind the
current peaks has not been fully understood.

Recently, a scattering theory based approach\cite{BSW,AB} has
successfully explained SGS in conventional ($s$-wave) junctions for
arbitrary transparency of the barrier: it agrees for one-channel
junctions with experiments\cite{vanP} without fitting parameters. In a
recent reformulation\cite{joh} of the theory it was shown that only
currents from multi-particle processes creating real excitations need
to be summed up, since non-physical currents, present in the original
formulation of the theory, cancels. This proves that the Pauli
exclusion principle is fulfilled.

In this paper we extend the reformulated scattering theory described
above to junctions of $d$-wave superconductors and provide a deeper
analysis of the SGS and the effects of MGS.

\section{EXPRESSION FOR THE CURRENT}

Considering transport in the $ab$-plane, we model the
$d_{\alpha_L}/d_{\alpha_R}$ junction ($\alpha_{L/R}$ is the
orientation angle of the left/right superconductor\cite{HLJW}) as
described in Ref.~\onlinecite{HLJW}.  This reference also provides a
detailed description of the method we use to solve the time-dependent
Bogoliubov-de Gennes (BdG) equation.  A quasiparticle incident on the
junction at energy $E$ undergoes multiple Andreev reflections and
builds up a scattering state with amplitudes at the sideband energies
$E_n=E+neV$ (V is the voltage; $n$ integer). It can be shown that the
probability current $I_n^p$ leaking out at sideband $E_n$ determines
the $n$-particle current $I_n$. At zero temperature we
have\cite{joh,unpub}
\begin{eqnarray}
&&I_{dc}(V) = \sigma_0 \sum_n nI_n(V),\hspace{1cm}
I_n(V)=\int_{-\pi/2}^{\pi/2} d\theta \cos\theta
\int_{-neV}^0 dE I_n^p(\theta,E),\nonumber\\
&&I_n^p(\theta,E) = \sum_{\alpha=\{l,r\}}
\left[(1-|a_0|^2)+(1-|{\bar a}_0|^2)|a_0r_{0-}|^2\right]
|G_{no}|^2\nonumber\\
&&\hspace{3cm}\times\left[(1-|a_n|^2)+
(1-|{\bar a}_n|^2)|a_nr_{n+}|^2\right],\label{eq:current}\\
&&G_{n0}  = \frac{t_{n0}}{(1-a_0{\bar a}_0r_{0-}r_{n0})
(1-a_n{\bar a}_nr_{n+}{\tilde r}_{n0})-
a_0{\bar a}_0a_n{\bar a}_nr_{0-}r_{n+}t_{n0}{\tilde t}_{n0}},\nonumber
\end{eqnarray}
where $\sigma_0=e k_F L_y/2\pi h$. Above, $a_0$ and $a_n$ are the
Andreev reflection amplitudes for the angle $\theta$ at energies $E_0$
and $E_n$ respectively (the barred amplitudes are calculated at
${\bar\theta}=\pi-\theta$); $r_{0-}$ and $r_{n+}$ describe reflections
from minus and plus infinity in energy space; $t_{n0}$, ${\tilde
  t}_{n0}$, $r_{n0}$, and ${\tilde r}_{n0}$ are the elements of the
scattering matrix describing the region between the injection point
and the exit point. The weight $n$ of $I_n$ appears because the
$n$-particle process involves an effective transfer of the charge $ne$
over the junction.

\section{RESULTS AND DISCUSSION}

\begin{figure}[t]
\centerline{\psfig{file=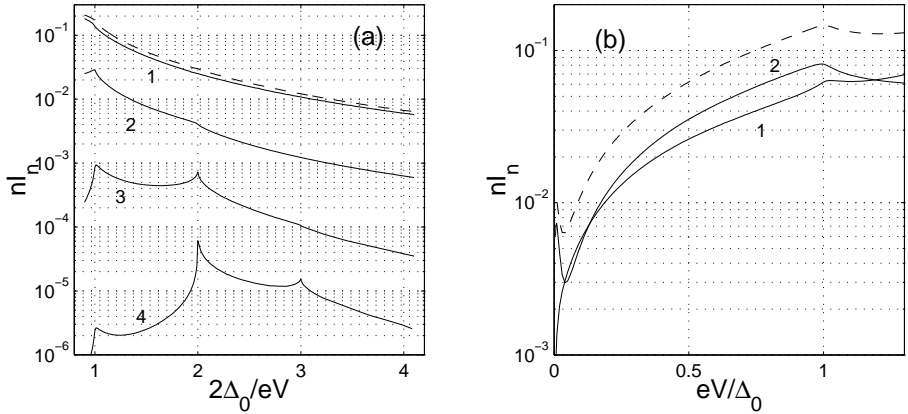,height=2.25in}}
\caption{{\small Contributions from the first four $n$-particle currents
    to SGS for the $d_0/d_0$ junction in (a), and the $d_0/d_{45}$
    junction in (b). The dashed lines are the total currents. In (b)
    broadening $\eta=0.01$ has been introduced revealing a current
    peak near zero bias. The angle averaged junction transparency is
    $D=0.026$ and we assume zero temperature. Note the inverse voltage
    scale in (a).}}
\label{fig1}
\end{figure}

Analyzing the current in Eq.~(\ref{eq:current}), we see that
resonances may appear in the propagator $G_{n0}$, which describes the
transmission process from energy $E$ to $E_n$. There are two types of
resonances: bare transmission resonances (appearing in $t_{n0}$) and
boundary resonances (due to the denominator). A bare transmission
resonance may appear when the trajectory hits a bound state at the IS
interface, so called de Gennes state.  In a short $d_0/d_0$ junction
(as in the short $s$-wave junction) the de Gennes states are located
at the gap edges, giving rise to the usual SGS at the voltages
$eV=2\Delta/n$. In the $s$-wave case\cite{BSW,joh} the SGS is due to
the combination of onset of the $n$-particle current and resonances in
the $n+1$-particle current (due to an overlap of a bare transmission
resonance and a boundary resonance) and the $n+2$-particle current
(two overlapping bare transmission resonances). For the $d_0/d_0$ case
[see Fig.~\ref{fig1}(a)] the physics behind the SGS is the same, but
for two reasons the structure is smeared and suppressed. First, there
are no real onsets (the $d$-wave gap has nodes), meaning that the
$1$-particle current background dominates at all voltages. In
addition, the resonances in the higher order currents are not sharp
because of angular averaging.

Rotating the right $d$-wave gap away from the $\alpha_R=0$
orientation, the de Gennes states are moved from the gap edges to zero
energy (MGS) for those angles where the gap changes sign after normal
reflection at the junction. The bare transmission resonance is then
moved and, in addition, the boundary resonances are lost for
quasiparticles injected from the right superconductor. This result in
a drastically changed SGS: for the $d_0/d_{45}$ junction [see
Fig.~\ref{fig1}(b)] the only surviving structure is at $eV=\Delta_0$
and it is mainly due to the $2$-particle current: an overlap between
the bare MGS resonance and boundary resonances (near the left gap
edges) produces the peak. This happens also in the $s/d_{45}$
junction\cite{LJSWH}. We introduce, on a phenomenological level,
inelastic scattering into the problem by adding a small imaginary part
$i\eta$ to the quasiparticle energy, which results in broadening of
all resonances.  For the $d_0/d_{45}$ junction, a small current peak
is then revealed near zero bias, as seen in Fig.~\ref{fig1}(b). The
peak is due to resonances in particle currents of order $n=2$ and
higher and was therefore not discussed in connection to the tunnel
limit calculations in Ref.~\onlinecite{BarSvi} and
\onlinecite{SamDat}.

\begin{figure}
\centerline{\psfig{file=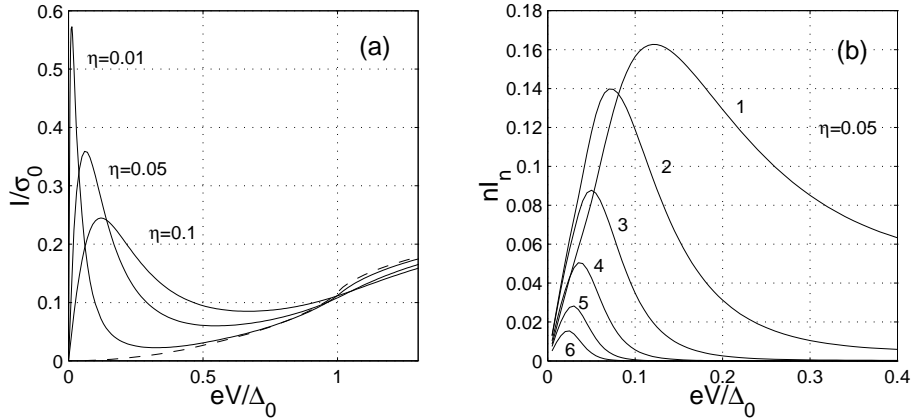,height=2.25in}}
\caption{{\small Introducing broadening reveals a current peak near zero bias
    in the $d_{45}/d_{45}$ junction. Decreasing $\eta$, sharpens the
    peak and moves it to lower voltage. In the limit $\eta\rightarrow
    0$ (dashed line) the peak becomes a delta spike at $V=0$. In (b)
    we show that particle currents of high order contribute to the
    peak.  The angle averaged junction transparency is $D=0.026$ and
    we assume zero temperature.}}
\label{fig2}
\end{figure}

When MGS are present on both sides of the junction (the
$d_{45}/d_{45}$ junction) boundary resonances can never appear.
Consequently current peaks are not seen in the
IV-characteristics\cite{Hurd}, leaving only an onset of the current at
$eV=\Delta_0$ (due to the bare MGS resonance). Again, introducing
broadening reveals a peak near zero bias, see Fig.~\ref{fig2}(a). In
Fig.~\ref{fig2}(b) we show that processes of many orders are in this
case contributing to the peak.

In recent experiments\cite{Alff_SS1,Alff_SS2} on bicrystal grain
boundary junctions of hole-doped cuprates a ZBCP was seen for all
orientations of the superconductors. No real SGS was seen, only a
gap-like structure at $eV=\Delta_0$. Our predicitions agree with the
experimental results, apart from that negative differential
conductance was not found in the experiments. In another
experiment\cite{Engel} weak SGS (altough not perfectly at
$eV=2\Delta_0/n$) was observed for edge junctions with the $d_0/d_0$
orientation, as we also report here.

\section{SUMMARY}
Dividing the current into $n$-particle currents, we have shown which
physical processes are giving rise to SGS and current peaks near zero
bias. For orientations where no MGS are present in the junction the
SGS is at $eV=2\Delta_0/n$ as in the $s$-wave case. When MGS are
present on at least one side of the junction SGS is lost and a current
peak appears near zero bias (if we include broadening into the
formalism). Currents of high orders are contributing to this peak.

\section*{ACKNOWLEDGMENTS}
It is a pleasure to thank V. S. Shumeiko for many useful discussions
on this work.  Economic support from NUTEK and NFR (Sweden) and NEDO
(Japan) is gratefully acknowledged.


\begin{thebibliography}{9}

\bibitem{Hu} C.-R. Hu, Phys. Rev. Lett. {\bf 72}, 1526 (1994).

\bibitem{Cov} M. Covington {\it et al.}, Phys. Rev. Lett. {\bf 79}, (1997), and references therein.

\bibitem{Alff_NS} L. Alff {\it et al}, Phys. Rev. B {\bf 55}, R14757 (1997), and references therein.

\bibitem{Hurd} M. Hurd, Phys. Rev. B {\bf 55}, R11993 (1997).

\bibitem{BarSvi} Yu. S. Barash, and A. A. Svidzinsky, Zh. Eksp. Teor. Fiz. {\bf 111}, 1120 (1997) [JETP {\bf 84}, 619 (1997)].

\bibitem{SamDat} M. P. Samanta, and S. Datta, Phys. Rev. B {\bf 57}, 10972 (1998).

\bibitem{Yoshida} N. Yoshida, Y. Tanaka, and S. Kashiwaya, preprints 1999.

\bibitem{BSW} E. N. Bratus', V. S. Shumeiko, and G. Wendin, Phys. Rev. Lett. {\bf 74}, 2110 (1995).

\bibitem{AB} D. Averin, and A. Bardas, Phys. Rev. Lett. {\bf 75}, 1831 (1995).

\bibitem{vanP} B. Ludoph, {\it et al}., submitted to Phys. Rev. B [cond-mat/9908223].

\bibitem{joh} G. Johansson {\it et al}., Superlattices and Microstructures, in press.

\bibitem{HLJW} M. Hurd {\it et al}., Phys. Rev. B {\bf 59}, 4412 (1999).

\bibitem{LJSWH} T. L\"ofwander {\it et al}., Superlattices and Microstructures, in press.

\bibitem{unpub} T. L\"ofwander, to be submitted for publication.

\bibitem{Alff_SS1} L. Alff {\it et al.}, Phys. Rev. B {\bf 58}, 11197 (1998).

\bibitem{Alff_SS2} L. Alff {\it et al.}, Eur. Phys. J. B {\bf 5}, 423 (1998).

\bibitem{Engel} A. Engelhardt, R. Dittmann, and A. I. Braginski, Phys. Rev. B {\bf 59}, 3815 (1999).

\end{thebibliography}
\end{document}